# Do gamma-ray bursts come from the Oort cloud?

*T. E. Clarke*[1], *Omer Blaes*[2], *and Scott Tremaine*
Canadian Institute for Theoretical Astrophysics
McLennan Labs, University of Toronto
60 St. George St., Toronto, Canada M5S 1A7

## Abstract

We examine the possibility that gamma-ray bursts arise from sources in the Oort comet cloud, basing most of our arguments on accepted models for the formation and spatial distribution of the cloud. We identify three severe problems with such models: (1) There is no known mechanism for producing bursts that can explain the observed burst rate and energetics without violating other observational constraints. (2) The bright source counts cannot be reconciled with standard models for the phase-space distribution of objects in the Oort cloud. (3) The observed isotropy of the available burst data is inconsistent with the expected angular distribution of sources in the Oort cloud. We therefore assert that Oort cloud models of gamma-ray bursts are extremely implausible.

## 1 Introduction

The BATSE experiment on board the *Compton Gamma Ray Observatory* (Meegan et al. 1992a) has demonstrated that cosmic gamma-ray bursts (GRBs) are isotropically distributed on the sky but inhomogeneously distributed in space. The GRB sources are therefore commonly believed to be situated in an extended halo around the Galaxy (at distances $D \sim 100\,\mathrm{kpc}$) or at cosmological distances ($D \sim 1$ Gpc). A third alternative, rarely discussed in the literature but generally acknowledged as a possibility, is that the GRB sources are much closer, at distances small compared to the thickness of the Galactic disk ($D \lesssim 100$ pc).

A firm lower limit to this distance is a few hundred AU; at smaller distances, the burst arrival wavefront would not be sufficiently planar for the burst direction to be localized within several arcminutes by the interplanetary network (K. Hurley 1993, private communication)[3]. Between 200 AU and 100 pc—a distance range of five orders of magnitude—there are two natural host sites for the GRBs.

The first is the local interstellar medium (e.g. Cox & Reynolds 1987). It is conceivable that GRBs might be associated with particular phases of the interstellar medium and hence their distance and angular distributions would reflect peculiarities of local geography. For example, the Sun appears to lie close to the centroid of a low-density HI cloud (the "Local

[1] Now at Dept. of Astronomy, McLennan Labs, University of Toronto, 60 St. George St., Toronto M5S 1A7, Canada
[2] Now at Dept. of Physics, University of California, Santa Barbara, CA 93106
[3] Thus the solar wind, which terminates at the heliopause ($D \sim 100\,\mathrm{AU}$), cannot play a role in GRBs.

Fluff"; number density $\sim 0.1\,\mathrm{cm}^{-3}$, temperature $\sim 10^4$ K, radius 3–5 pc); if GRBs were confined for some reason to such clouds, then their inhomogeneous spatial distribution and isotropy could perhaps be explained in terms of the geometry of the Local Fluff.

The second natural site to consider, and the one we shall focus on in this paper, is the Oort (1950) comet cloud, a roughly spherical distribution of $N \sim 10^{12}$ comets at a typical distance $D \sim 10^4$ AU. The typical cometary mass and radius are $m_c \sim 10^{17}$ g and $R \sim 3\,\mathrm{km}$ (for reviews see Weissman 1990 and Fernández & Ip 1991).

Much of our analysis is based on the following standard model of the formation of the Oort cloud (e.g. Duncan, Quinn & Tremaine 1987, hereafter DQT). As the protoplanetary disk cools, solid bodies ("planetesimals") are formed by gravitational instability. The size distribution of planetesimals evolves as they collide, accrete and fragment. A substantial fraction is eventually concentrated in a few large objects, which become the cores of the giant planets. Many of the residual planetesimals—which we identify with comets—are gravitationally scattered by the giant planets; through this process, the distribution of comets diffuses to higher and higher energies (larger semimajor axes), but the perihelia $q$ of the comets remain within the planetary region ($q \lesssim 30\,\mathrm{AU}$) since this is the only source of perturbations. The masses of Jupiter and Saturn are sufficiently large that most comets under their influence ($q \lesssim 15\,\mathrm{AU}$) are scattered onto escape orbits and lost from the solar system, but the evolution of comets scattered by Uranus and Neptune ($q \gtrsim 15\,\mathrm{AU}$) is more complex. When the cometary semimajor axis exceeds a few thousand AU, the torque from the Galactic tidal field increases the angular momentum (and hence the perihelion) enough so that the comet no longer enters the planetary region and hence is immune to further energy diffusion. The tide continues to change the angular momentum and eccentricity of the comet orbits, thereby generating a roughly spherical comet cloud (the Oort cloud) surrounding the solar system. Comets slowly leak out of the cloud as the result of perturbations from passing stars or because their perihelia precess back into the planetary region, but most comets in the cloud survive for the lifetime of the solar system.

Comets are seen by reflected sunlight and hence are only visible if their perihelion distance $q \lesssim 2\,\mathrm{AU}$; thus Oort cloud comets are only detectable if their angular momenta are very small (eccentricity exceeding 0.9999). Most long-period comets (those with orbital periods exceeding 200 y) arise from the outer parts of the Oort cloud, semimajor axes $a \gtrsim 2 \times 10^4$ AU, as the Galactic tide and perturbations from passing stars sweep a steady flux of new comets into low angular momentum orbits. Oort cloud comets from smaller semimajor axes are not seen: external perturbations are less effective on smaller orbits, so that these comets are removed already from the cloud by Jupiter and Saturn long before they would be visible from Earth. Comets from this "inner" cloud are only visible during rare comet showers lasting a few My, caused by an unusually strong encounter with a passing star (Hills 1981). Simulations of the scattering process that formed the Oort cloud (see DQT) suggest that the inner cloud contains 80% of all Oort cloud comets and occupies the range of semimajor axes $3 \times 10^3$ AU $\lesssim a \lesssim 2 \times 10^4$ AU.

Of course, any exotic bodies (e.g. small black holes) that are present in the protoplanetary disk would populate the Oort cloud by the same process.

Maoz (1993) has compared the angular distribution of 260 GRBs observed by BATSE with the distribution of aphelion directions of known long-period comets. He found that it is unlikely that the two distributions agree, although $\gtrsim 800$ burst locations would be needed to establish a difference between the distributions at the 1% level. Unfortunately, this test is meaningless for at least two reasons. First, if GRBs are associated with comets in the Oort cloud they must come from the inner cloud, since it is both closer and more populous than the "active" Oort cloud that is the source of observed comets; thus the observed long-period comets and the hypothesized GRB sources are on orbits that differ in semimajor axis by an order of magnitude and in perihelion by more than three orders of magnitude. Second, the angular distribution of observed comets is likely to depend strongly on the parameters of the last few close stellar encounters, since the typical infall time of a comet from the active Oort cloud is comparable to the mean time between encounters of passing stars with the cloud (both are a few My).

The number of GRBs as a function of peak count rate can also be compared with the predictions of Oort cloud models. Such a comparison was also carried out by Maoz (1993) but again is unrealistic since it is based on the observed long-period comets. Horack et al. (1993) have also compared Oort cloud models with the BATSE data. In particular they find no significant concentration of bursts towards the ecliptic plane or towards the (time-dependent) position of the Sun on the sky. They also compared the peak count rate distribution with predictions of Oort cloud models; we discuss their results in §3.

We argue in this paper that models in which GRBs arise in the Oort cloud face several crippling difficulties. In §2 we show that the overall energetics and burst rate requirements are extremely difficult to satisfy with Oort cloud models. In §§3 and 4 we argue that the intensity and angular distributions of GRBs are incompatible with the probable properties of the Oort cloud. We summarize our conclusions in § 5.

## 2 Energetics and burst rate

Even in this relatively nearby region of space, the energy requirements for bursts are high. The energy of a source at distance $D$ that produces a burst with observed fluence $F$ is

$$E_b = 3 \times 10^{28} f_b \left(\frac{D}{10^4\,\mathrm{AU}}\right)^2 \left(\frac{F}{10^{-7}\mathrm{ergs\,cm^{-2}}}\right)\ \mathrm{ergs}, \tag{1}$$

where $f_b$ is a beaming factor that is unity if the emission is isotropic.

A natural starting point in examining the plausibility of Oort cloud models is to ask whether comet-comet collisions can make GRBs. There are at least three strong reasons why they cannot: (1) The maximum kinetic energy of a bound comet is $GM_\odot m_c/D = 9 \times 10^{25}(m_c/10^{17}\mathrm{g})(10^4\,\mathrm{AU}/D)$ ergs. This is far smaller than the typical burst energy $E_b$, assuming that the faintest observed bursts ($F \sim 10^{-7}\mathrm{ergs\,cm^{-2}}$) come from the outer parts of the Oort cloud ($D \gtrsim 10^4$ AU). (2) It is difficult to understand why the collision energy is released in gamma rays: the temperature of a black body of radius $R$ that emits energy $E_b$ in a burst duration $\Delta t$ (typically 10 s; see, e.g., Higdon & Lingenfelter 1990) is $T = [E_b/(4\pi R^2 \sigma \Delta t)]^{1/4} = 6 \times 10^4\ \mathrm{K}\,(E_b/10^{28}\ \mathrm{ergs})^{1/4}(3\,\mathrm{km}/R)^{1/2}(10\ \mathrm{sec}/\Delta t)^{1/4}$, far too low to produce significant gamma ray emission. (3) The collision rate is much too low: the mean time between collisions in a population of $N$ objects of radius $R$, spread uniformly over a spherical volume of radius $D$ and moving at typical speed $v = (GM_\odot/D)^{1/2} = 0.3\,\mathrm{km\,s^{-1}}(10^4\ \mathrm{AU}/D)^{1/2}$ is

$$\sim 0.2D^3/(N^2R^2v) \sim 10^4\ \mathrm{y}(D/10^4\ \mathrm{AU})^3(10^{12}/N)^2(3\ \mathrm{km}/R)^2(0.3\,\mathrm{km\,s^{-1}}/v).$$

For comparison the inferred burst rate is $\sim 800/f_b\ \mathrm{y}^{-1}$ (Meegan et al. 1992a), a factor of $10^7$ higher.

A more exotic possibility is that the Oort cloud also contains small black holes, which produce GRBs when they collide with comets (Bickert & Greiner 1993). Let us suppose that the Oort cloud contains $N_h$ black holes of mass $m_h$, with a spatial distribution similar to that of the comets. Since we have seen that the comet-comet collision rate is too low by a factor of $10^7$ to produce the observed burst rate, we must have $N_h \sim 10^7 N \sim 10^{19}$ (assuming the cross-section is similar to the comet size). This model also has serious flaws: (1) The black hole mass must exceed $m_h = 5 \times 10^{14}$ g or else the holes would have evaporated by Hawking radiation in the lifetime of the universe; thus the total mass in holes $N_h m_h$ exceeds a solar mass, far larger than the dynamical limits on unseen mass in the outer solar system (Tremaine 1989). (2) This contradiction is strengthened considerably by the plausible requirement that the burst energy radiated over a typical burst duration should not exceed the Eddington luminosity, which requires $m_h \gtrsim 10^{22}\ \mathrm{g}(E_b/10^{28}\ \mathrm{ergs})(10\ \mathrm{s}/\Delta t)$. (3) If the spatial distribution of the black holes is similar to that of the comets, then the black holes, like comets, will occasionally collide with the Earth. Assuming that the black holes are uniformly distributed on the energy hypersurface in phase space (which is at least true in the outer, active, Oort cloud), and neglecting repeated encounters by black holes that are captured onto tightly bound orbits by planetary perturbations (which enhance the collision rate), the mean interval between collisions with the Earth is $2 \times 10^6\ \mathrm{y}\,(D/10^4\ \mathrm{AU})^{5/2}(10^{12}/N_h)$, where we have identified $D$ with the semimajor axis. If, as the model requires, $N_h \sim 10^{19}$, the interval between collisions is less than a year. The energy released in a collision between a black hole and the Earth is difficult to estimate but—if this model is correct—can hardly be less than the energy released in a GRB, i.e. of order $10^{28}$ ergs (eq. 1) or over $10^5$ Megatons of TNT, which could not escape detection.

If the hole mass is sufficiently large, $m_h \gtrsim M_\odot R/D \simeq 4\times10^{21}\ \mathrm{g}(R/3\ \mathrm{km})(10^4\ \mathrm{AU}/D)$, then gravitational focusing enhances the collision cross-section to $\sim 3Gm_hR/v^2$ (and possibly somewhat more if the comet is tidally disrupted after a near-miss). If there are $N$ comets, then the total collision rate would be

$$\sim NN_hGm_hR/(vD^3) = 10^{-5}\mathrm{y}^{-1}(N/10^{12})(N_hm_h/M_\odot)(R/3\ \mathrm{km})(10^4\ \mathrm{AU}/D)^{5/2}.$$

This is still far lower than the observed burst rate for the typical comet mass and likely Oort cloud population, even if the Sun is surrounded by a solar mass of black holes. Higher collision rates are possible for plausible populations of smaller bodies, so we now examine this possibility. If the entire rest mass energy of the comet is converted into gamma rays, the required burst energy of $10^{28}$ ergs could be produced by the collision of a $10^7$ g object with a black hole. The corresponding radius is $R \sim 100$ cm; for a total population of $N \sim 10^{25}$ such "microcomets" and a total black hole mass $N_h m_h \sim 0.1 M_\odot$ (so that the total microcomet mass and black hole mass are close to the upper limit allowed by dynamical constraints; see Tremaine 1989), we obtain a collision rate $\gtrsim 10^3\ \mathrm{y}^{-1}$ as required by the GRB rate. The collision rate of microcomets with the Earth is then $\sim 10^7\ \mathrm{y}^{-1}$ (using the formula in the preceding paragraph). This model has two interesting features: first, the mass of the microcomet is so small that the material accreting onto the black hole can be optically thin, consistent with the nonthermal spectrum of the burst; second, Frank et al. (1986) have independently argued that microcomets with similar mass ($10^8$ g) impact the Earth at a similar rate ($10^7\ \mathrm{y}^{-1}$), based on observations of transient dark spots in the ultraviolet emission of the Earth's upper atmosphere. Despite this interesting coincidence, the required population of microcomets is incompatible with observations: a $10^7$ g object impacting the Earth or Moon at escape speed from the solar system releases an energy of approximately 1 kiloton of TNT, and the estimated rate of such collisions with the Earth is only $10\ \mathrm{y}^{-1}$ (Shoemaker 1983), a million times too small (see also Dessler 1991 for a comprehensive criticism of Frank et al.'s microcomet hypothesis; most of Dessler's criticisms also apply to the model described here).

The energetics are also a problem for models in which GRBs arise in the Local Fluff: the total magnetic field or cosmic ray energy in an average spherical volume of radius 60 km (constrained by the short variability timescale of at least one burst, see Bhat et al. 1992) in the Local Fluff, e.g. $E_{mag} \sim 10^9 (B/5\mu\mathrm{G})^2$ ergs, is far smaller than the burst energy.

## 3 Intensity Distribution

The best statistics on the brightness distribution of strong bursts are given by the *Pioneer Venus Orbiter* (PVO) data set. Here the differential source counts as a function of peak photon count rate $C_{max}$ are proportional to $C_{max}^{\alpha}$, with $\alpha = -2.42 \pm 0.07$, consistent with a homogeneous spatial distribution of sources, for which $\alpha = -\frac{5}{2}$ (Chuang et al. 1992). We now show that the distribution of comets is not expected to have a homogeneous central core (cf. Paczyński 1993), and thus the bright source counts produce severe constraints on Oort cloud models.

For example, let us make the modest assumptions that the burst rate is proportional to the local density of comets, and that the phase-space density $f$ of the comets is constant on energy hypersurfaces, i.e. $f$ depends only on the energy per unit mass $E$. The spatial density is then

$$n(r) = \int f\, d\mathbf{v} = 4\pi \int_{-GM_\odot/r}^{0} dE \left(2E + 2\frac{GM_\odot}{r}\right)^{1/2} f(E), \tag{2}$$

where $r$ is the distance from the Sun (or the Earth at the distance scales of interest). Simple manipulations yield

$$\frac{d(nr^{1/2})}{dr} = \frac{4\pi}{r^{1/2}} \int_{-GM_\odot/r}^{0} E\, dE \left(2E + 2\frac{GM_\odot}{r}\right)^{-1/2} f(E). \tag{3}$$

Since the right-hand side is negative, the number density $n(r)$ must increase at least as fast as $r^{-1/2}$ as $r \to 0$. In particular, provided $f(E)$ vanishes sufficiently rapidly as $E \to -\infty$, then $n \propto r^{-1/2}$ as $r \to 0$. A homogeneous central core in which $n(r)$ is constant as $r \to 0$ is impossible.

The implications of this conclusion are as follows. Consider a population of sources with density distribution given by $n(r) = Ar^{-\beta}$, where $\beta < 3$ and $A$ is a constant. The luminosity (or strictly speaking, peak photon emission rate) function $\Phi(L)$ is assumed to be independent of position. The cumulative source counts are given by

$$N(> C_{\mathbf{max}}) = 4\pi \int_0^\infty dL\Phi(L) \int_0^{(L/4\pi C_{\mathbf{max}})^{1/2}} dr r^2 n(r) \tag{4}$$

which is $\propto C_{\mathbf{max}}^{(\beta-3)/2}$ provided $\int dL\Phi(L)L^{3-\beta}$ converges. Even if the power-law density relation is cut off at some maximum radius $r_{\mathbf{max}}$, then

$$\frac{d}{dC_{\mathbf{max}}}\left[C_{\mathbf{max}}^{(3-\beta)/2} N(> C_{\mathbf{max}})\right] = 2\pi A r_{\mathbf{max}}^{3-\beta} C_{\mathbf{max}}^{(1-\beta)/2} \int_{4\pi C_{\mathbf{max}} r_{\mathbf{max}}^2}^\infty dL\Phi(L). \tag{5}$$

Since the right hand side is positive, the logarithmic slope of the cumulative counts can nowhere be steeper than $(\beta - 3)/2$. Inserting $\beta > \frac{1}{2}$ from the discussion in the preceding paragraph, we find that the cumulative bright source counts must have a logarithmic slope $> -1.25$, a value $2.4\sigma$ away from the PVO result $-1.42 \pm 0.07$. Continued accumulation of counts by BATSE may sharpen this discrepancy.

To confirm this conclusion, we have performed a Monte Carlo simulation of bursts arising from objects on orbits with the semimajor axis distribution expected for the Oort cloud (taken from Fig. 8 of DQT) and a uniform distribution of squared eccentricities (which corresponds to a phase-space density depending only on energy). Bursts were either chosen to be standard candles or to have luminosities drawn from power law distributions with a variety of exponents. The steepest logarithmic slope of the cumulative bright source counts that we found was $\simeq -1.2$, formally inconsistent (at $3.1\sigma$) with the PVO value. The discrepancy between the Monte Carlo slope and the theoretical value of $-1.25$ arises because the density distribution approaches its asymptotic behavior $n \propto r^{-0.5}$ only slowly; at the smallest radii in our simulations ($250\,\mathrm{AU} \lesssim r \lesssim 1000\,\mathrm{AU}$) the density is still $\propto r^{-0.65}$.

How realistic is the assumption that the phase-space density of comets depends only on energy? Simulations of the formation of the Oort cloud by scattering comets from the protoplanetary disk (DQT) show that for semimajor axes exceeding $\sim 2000\,\mathrm{AU}$ (a region that contains most of the comets in the Oort cloud) the mean-square eccentricity is near the value $\frac{1}{2}$ that is expected if the phase-space density depends only on energy. At smaller radii the eccentricity is larger (the orbits are more nearly radial), which exacerbates the discrepancy with the observed counts, since the density on near-radial orbits increases even more rapidly toward the center.

It is of course possible to set up *ad hoc* orbit distributions (e.g. purely circular orbits around 500 AU) that could generate an intensity distribution that is consistent with the data. However, there is no known formation process that would establish such an orbit distribution.

We have also assumed that the burst rate is proportional to the comet density. Models in which the burst rate is more sensitive to the source density (e.g. collision models, in which the burst rate is proportional to the square of the density), or in which the rate increases with the velocity of the sources, will have even steeper concentration of bursts towards small radii, and are therefore even more inconsistent with the data.

In contrast to our conclusions, Horack et al. (1993) claim that the intensity distribution expected from Oort cloud models is consistent with observations. Their chosen density distribution varies as $n(r) \propto r^{-\beta}$, $\beta = 1.5$, at small distances and is zero inside a cutoff distance $R_{\mathbf{min}}$. In the absence of a cutoff, we would expect their cumulative bright source counts to have a logarithmic slope of $(\beta - 3)/2 = 0.75$, rather consistent with the *faint* BATSE counts and in agreement with their results.[4] These authors succeed in obtaining agreement with the steeper bright source counts by the use of the *ad hoc* finite cutoff radius $R_{\mathbf{min}}$ inside of which there are no bursts. Thus they implicitly assume that the phase-space density is not constant on the energy hypersurface. Their model requires either an unknown formation mechanism that preferentially places comets on near-circular but randomly oriented orbits, or some unknown process that prohibits bursts when a comet comes too close to the Sun.

## 4 Sky distribution

The standard model for the formation of the Oort cloud is based on the action of the Galactic tide, which lifts cometary perihelia out of the planetary region, thereby protecting them from repeated planetary perturbations. The continued action of tidal forces will spread the comet orbits into a steady-state distribution whose angular shape depends on the angle between the ecliptic and the Galactic plane, $I_e = 60.2°$, and which in general is not isotropic. If GRBs come from the Oort cloud, their present angular distribution should reflect this formation process in the following way:

(i) *Small semimajor axis* ($a \lesssim 3 \times 10^3$ AU): In this region the tidal torque has not had a significant influence on the overall shape of the orbit over the lifetime of the solar system. Thus the comets, and hence the GRBs, will still be strongly concentrated to the ecliptic. This distribution is excluded by the BATSE observations, which show no concentration to the ecliptic (Horack et al. 1993). The upper limit to the semimajor axis range of this region is obtained by equating the typical time for the perihelion to precess from near-zero to its maximum value ($0.7 \times 10^9$ y$(10^4$ AU$/a)^{3/2}$; see DQT) with the age of the solar system.

(ii) *Intermediate semimajor axis* ($3 \times 10^3$ AU $\lesssim a \lesssim 3 \times 10^4$ AU): In this region the precession time is less than the age of the solar system, so that a steady-state distribution of orbits has been established under the influence of the Galactic tide.

[4] Although Horack et al. (1993) only considered standard candles, this slope is to be expected for general luminosity functions as well.

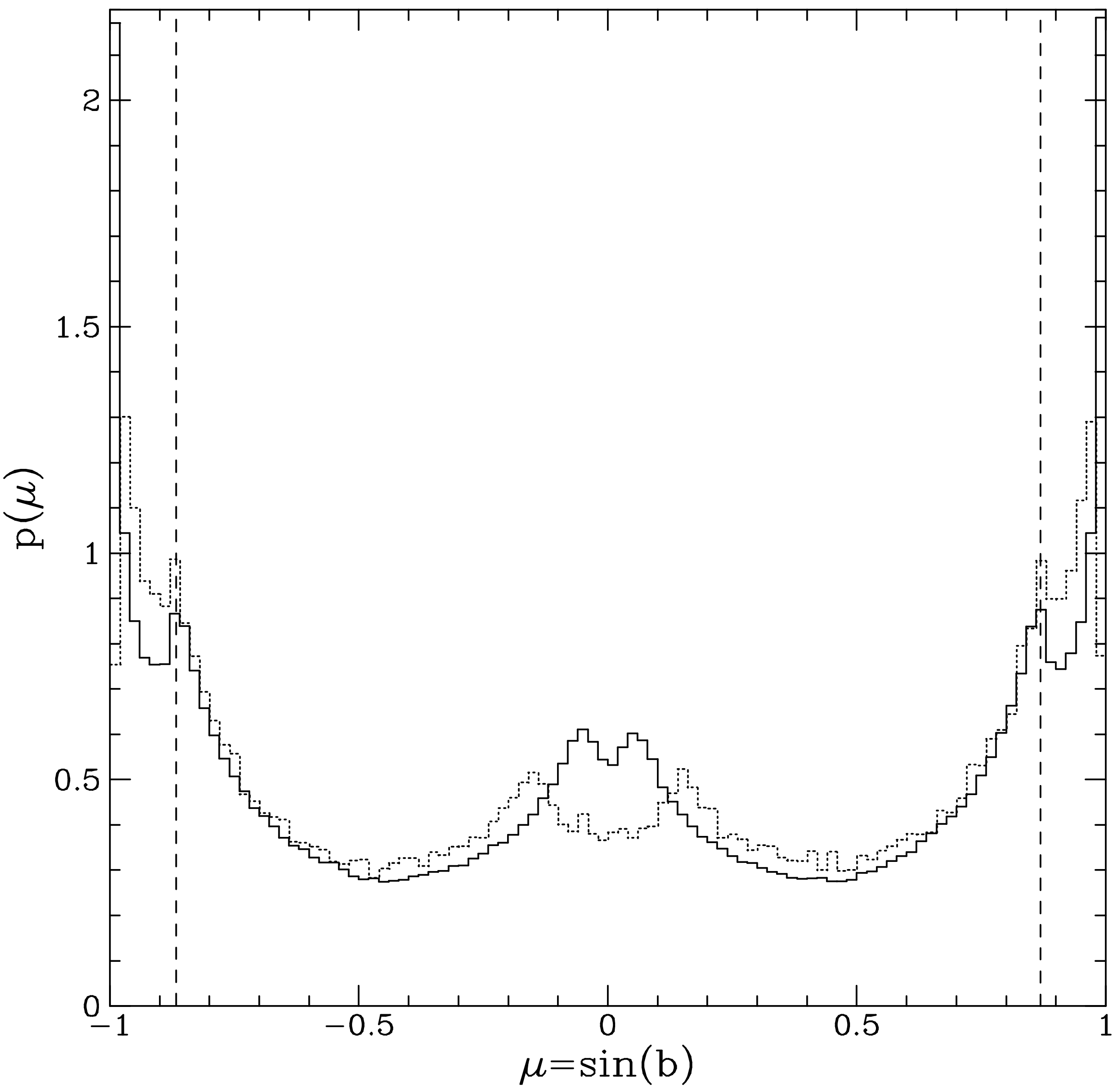

**1.** Probability distribution of burst Galactic latitudes from Monte Carlo simulation of $10^6$ orbits with initial eccentricity $e_i = 0.997$ (solid line). Also shown are the results of a somewhat noisier simulation of $10^5$ orbits with $e_i = 0.95$ (short dashed line). The vertical long dashed lines are at $b = \pm I_e$, where $I_e = 60.2^\circ$ is the angle between the ecliptic and the Galactic plane.

The upper limit to the semimajor axis is set by the condition that the isotropization time from perturbations by passing stars ($1.5 \times 10^{10}$ y$(10^4\,\mathrm{AU}/a)$, from Heisler & Tremaine 1986, eq. 33) must exceed the age of the solar system.

The angular distribution expected in this region depends only on Galactic latitude $b$. The expected distribution is computed in the Appendix and shown in Figure 1. There is a concentration towards the Galactic poles, which is reflected in the statistic $\langle \sin^2 b \rangle = 0.453$. The standard deviation in this statistic for large numbers of bursts $N_b$ (calculated by Monte Carlo simulation and the central limit theorem) is $0.32 N_b^{-1/2}$. Thus the angular distribution in region (ii) is highly inconsistent (at $9\sigma$) with the measured value $\langle \sin^2 b \rangle = 0.316$ for 447 BATSE bursts (Meegan et al. 1992b; we have not corrected for the nonuniform BATSE sky exposure but this correction would not alter our conclusion). We have also compared our angular distribution with the distribution of 260 public BATSE burst positions (Fishman et al. 1992) using a Kolmogorov-Smirnov test, and found that the probability that they are consistent is only $\simeq 3 \times 10^{-4}$, whether or not sky exposure corrections are included.

(iii) *Large semimajor axis* ($a \gtrsim 3 \times 10^4$ AU): In this region the comet distribution has been isotropized by stellar perturbations. While an isotropic distribution is consistent with the BATSE results, this region cannot be the source of the GRBs: region (ii) is both closer to the Sun and more populous than region (iii) (DQT estimate that region (iii) contains only 10% of Oort cloud comets) and hence should provide many more bright bursts if the GRBs do come from the Oort cloud.

We conclude that the distribution of GRBs on the sky that is expected if they are distributed in proportion to comets in the Oort cloud is incompatible with observations.

## 5 Conclusions

Comets or more exotic bodies that are present in the outer parts of the protoplanetary disk can be ejected into the Oort cloud by planetary perturbations. We have investigated the possibility that GRBs arise from sources in the Oort cloud, using three independent approaches: (1) Assuming that the burst rate is proportional to the density of these hypothetical sources, we have found that the expected bright source counts are inconsistent with the PVO data by at least $2.4\sigma$. The discrepancy is even larger if the burst rate is proportional to the square of the density. (2) The expected angular distribution of bursts is inconsistent with the BATSE data by $9\sigma$. (3) We have failed to find any plausible mechanism that generates GRBs at the required rate from any hypothetical source without violating other observational constraints, in particular, the observed rate of impacts on the Earth and Moon. The combination of these three arguments implies that Oort cloud models for gamma-ray bursts are extremely implausible.

Our arguments do not exclude the possibility that GRBs arise from a population of sources placed on *ad hoc* orbits that are outside the planetary system but inside the Oort cloud, and that avoid the inner planetary system (e.g. nearly circular orbits at 500 AU), but there is no independent evidence for such a population, nor is there any plausible way to populate such orbits during the formation of the solar system.

We thank K. Hurley, B. Paczyński, and D. Syer for useful discussions. This research was supported by NSERC.

## Appendix

We wish to determine the steady-state distribution of orbits that is established under the influence of the Galactic tide. This distribution should be present at semimajor axes between about $3\times10^3$ and $3\times10^4$ AU [region (ii) of §4], though its accuracy is questionable near the limits of this range.

Heisler & Tremaine (1986) have analyzed the effect of the Galactic tide on comet orbits. Because the horizontal tidal forces (i.e. forces in the Galactic plane) are much smaller than the vertical forces, the gravitational potential is well approximated by $-GM_\odot/r+\frac{1}{2}\nu^2z^2$, where $\nu$ is the vertical frequency of a test particle in the field of the Galaxy. Since the tidal perturbations are weak, Heisler and Tremaine average the perturbing Hamiltonian over the short orbital period of the comet $(10^6(a/10^4\ \mathrm{AU})^{3/2}$ y; "short" is in comparison to the precession time). They identify three integrals of motion in the averaged Hamiltonian[5]: the semimajor axis $a$, a dimensionless angular momentum $K_z\equiv(1-e^2)^{1/2}|\cos I|$, and a dimensionless energy

$$C\equiv 1-e^2+5e^2\sin^2 I\sin^2\omega = K^2+5(1-K^2)(1-K_z^2/K^2)\sin^2\omega, \qquad \text{(A-1)}$$

where $e$ is the eccentricity, $K\equiv(1-e^2)^{1/2}$, $I$ is the inclination of the orbit with respect to the Galactic plane, and $\omega$ is the argument of perihelion (the angle in the orbital plane from the horizontal to the perihelion).

Comets diffuse in semimajor axis under the influence of planetary perturbations until they reach a semimajor axis $a_f$ at which the Galactic tidal field removes their perihelia from the planetary region. At this point the inclination $I_i$ equals the inclination of the ecliptic, $I_e=60.2°$ (which has not changed significantly over the lifetime of the solar system). The perihelion distance $q\simeq30$ AU (since small bodies are most efficiently scattered to the Oort cloud by Uranus and Neptune), which implies an eccentricity $e_i=0.997$ if $a_f\simeq10^4$ AU (our results are not sensitive to the choice of initial eccentricity). Thus all of our comets have a single near-zero value of $K_z$, which is conserved after they enter the Oort cloud. The initial value of $\omega$ is randomly distributed between 0 and $2\pi$, as there is no preferred azimuth in the ecliptic plane. The distribution of initial $\omega$ determines the distribution of the constant of motion $C$: the fraction of comets with $C$ in the interval $[C, C+dC]$ is $p(C)dC$, where

$$p(C)=\frac{1}{\pi(C-C_{\mathbf{min}})^{1/2}(C_{\mathbf{max}}-C)^{1/2}}, \qquad \text{(A-2)}$$

with $C_{\mathbf{min}}\equiv1-e_i^2$ and $C_{\mathbf{max}}\equiv1-e_i^2+5e_i^2\sin^2 I_i$.

The position of a comet in phase space can be described by Delaunay's canonically conjugate momenta and coordinates, $(L,\ell)$, $(KL,\omega)$, $(K_zL,\Omega)$, where $L=(GM_\odot a)^{1/2}$, $\ell$ is the mean anomaly, and $\Omega$ is the longitude of the ascending node. The steady-state distribution of comets is described by the phase-space density $f$, defined so that $f dLdGdHd\ell dgdh$ is the fraction of comets in a small volume of phase space. Jeans' theorem implies that $f$ depends only on the three integrals of motion, $f=f(a,K_z,C)$. Since the semimajor axis $a$ determines the distance rather than the angular position and hence is of little interest, and since $K_z$ is the same for all comets, we may ignore the dependence on $a$ and $K_z$ and write $f = f(C)$. The angles $\ell$ and $\Omega$ are uniformly distributed between 0 and $2\pi$ since they are ignorable in the averaged Hamiltonian (the uniform distribution of $\Omega$ implies that the sky distribution will depend only on Galactic latitude, not longitude). Thus the fraction of comets between $C$ and $C + dC$ is

$$p(C) = 4\pi^2 f(C) \int dG dg \delta[C - C(G,g)] = 4\pi^2 F(C) \int dK d\omega \delta[C - C(K,\omega)], \qquad \text{(A-3)}$$

where $F = Lf$ and $C(K,\omega)$ is defined in equation (A-1). Our goal is to determine $F(C)$ by inverting equation (A-3), using equation (A-2) for $p(C)$.

[5] The analytic solution of the equations of motion in the averaged Hamiltonian is given by Matese & Whitman 1989, but only the integrals are needed for the present discussion.

This inversion is easily done using a Monte Carlo simulation. We bin the interval from $C_{\mathbf{min}}$ to $C_{\mathbf{max}}$ into $p$ bins and distribute $N_b$ bursts among these bins using equation (A-2), so that there are $N_j$ bursts in the bin centered on $C_j$. Each bin in $C$ corresponds to a region on the $(K,\omega)$ plane in which the density of comets is approximately constant. We pick a random position on the $(K,\omega)$ plane by choosing random pairs of values of $K$ (between $K_z$ and 1) and $\omega$ (between 0 and $2\pi$). For each pair we calculate the corresponding value of $C$; if this value lies in bin $i$ we set $N_j \leftarrow N_j - 1$ and add the pair $(K,\omega)$ to our list of orbital elements unless $N_j \leq 0$ in which case the pair is discarded. We continue to choose pairs of orbital elements in this manner until all the bins are empty; this process gives us a set of $N_b$ element pairs $(K,\omega)$ whose distribution satisfies equations (A-2) and (A-3). Each pair of elements is then supplemented with a random mean anomaly $\ell$; the triplets $(K,\omega,\ell)$ then determine the Galactic latitude $b$ of each comet.

The latitude distribution resulting from a simulation with $N_b = 10^6$ is shown in Figure 1. An isotropic distribution would have a constant value of 0.5 on this plot. The dashed line shows a simulation with a smaller initial eccentricity, $e_i = 0.95$; while noisier ($N_b = 10^5$ only in this case), it verifies that the latitude distribution is not strongly dependent on the initial eccentricity. The principal features of the plot can be understood analytically. In particular, the cusp near $\sin b = 0.87$ arises because orbits linger at latitudes near the inclination of the ecliptic, $I_e = 60.2°$.